\documentclass[twocolumn]{jpsj2} 
\title{Nonlinear Transport and Current Fluctuation in an AB Ring with a Quantum Dot}

\author{\textsc{Junko Takahashi}\thanks{E-mail address: junchi@ruri.waseda.jp} and \textsc{Shuichi Tasaki}}
\inst{Department of Applied Physics, School of Science and Engineering, Waseda University, Tokyo 169-8555}

\abst{Nonequilibrium steady states are explicitly constructed for a noninteracting electron
model of an Aharonov-Bohm (AB) ring with a quantum dot (QD) with the aid of asymptotic fields.
The Fano line shapes and AB oscillations are shown to strongly depend on the bias voltage.
Current fluctuations are studied as well.
}

\kword{Fano effect, quantum dot, AB ring, current fluctuation, nonequilibrium steady states}

\begin{document}
\maketitle

\section{Introduction} 
For an electron transport in an Aharonov-Bohm (AB) ring with a quantum dot (QD),
a coupling between the discrete states and continuum causes an asymmetry
of the resonance peak in the conductance\cite{Iye},
the so-called Fano resonance \cite{FANO}.
As well known, when one varies a magnetic field piercing into the AB ring,
the conductance oscillates (AB oscillation). 
And the phase of the AB oscillation changes by $\pi $ from one side of the Fano
resonance to the other\cite{Yacoby}. 
It is then interesting to see how these features appear in a far-from-equilibrium regime.

Recently, the method of C$^*$-algebra has been widely applied to 
rigorous investigations of nonequilibrium steady states (NESS) of infinitely
extended quantum systems\cite{Ojima,Ruelle,HoAraki,Jaksic,ST,Frolich}.
The method was originally developed to deal with equilibrium statistical mechanics of
such systems\cite{Bratteli,RuelleEq}.
Fr\"ohlich, Merkli and Ueltschi\cite{Frolich} have shown the existence of NESS  
and its characterization for junction systems, which consist of several free-fermionic 
reservoirs mutually interacting via bounded interactions.
Their results cover the AB ring with a QD in case that the electrodes are free fermion systems,
but explicit expressions of observables at NESS are not given.

In this article, exact NESS is constructed explicitly for a noninteracting-electron model of 
the AB ring with a QD. We then examine the bias voltage dependence of the Fano resonance shapes 
and the AB oscillation. Near the Fano peak, AB oscillation patterns are shown to be strongly
affected by the bias voltage.
This feature is shown to imply nonlinear conduction which is sensitive to the magnetic field.
Although we consider a noninteracting system, we believe that the Coulomb repulsion does not 
affect the essential features of the bias-voltage-dependence of the AB oscillation. 
The current fluctuation is investigated as well and we find that the interference effect appears 
in the fluctuation even when it cannot be seen in the average current.
  
\section{System and Nonequilibrium Steady State}

The system consists of two-dimensional
electrodes and a quantum dot interacting with each other,
which is schematically shown in Fig.1 \cite{Walter}. 
Only the dynamics of electrons is considered and the electron-electron interaction
is neglected. Then, its Hamiltonian is given by  \break
\indent
$H=H_L+H_R+H_D+H_{T}+H_{LR} \ ,$ \hfil \break
where $H_{\lambda }=\int dk\omega _{k\lambda }a_{k\sigma \lambda }^{\dagger }a_{k\sigma \lambda }$
($\lambda=L \ (R)$) is the energy of the left (right) electrode and $a_{k\sigma \lambda}$ ($\lambda=L \ (R)$)
is the annihilation operator
of an electron with wave number $k$ and spin $\sigma$ at the corresponding electrode. Throughout this article,
the repeated spin index means the summation over it. 
The bias voltage $V$ is taken into account as the
difference of the one-particle energies: $\omega _{kR}=\omega _{kL}-eV$.
The second term $H_D=\epsilon _{g }c_{\sigma }^{\dagger }c_{\sigma }$ stands for the 
energy of the QD, where $\epsilon_g$ is the dot level and $c_\sigma$ is the annihilation operator
of the electron with spin $\sigma$ at the dot. 
The electron transfer through the QD is described by a tunneling
Hamiltonian \hfil \break
\indent $H_{T}=\int dk \{u_{kL} a_{k\sigma L}^{\dagger }c_{\sigma }+u_{kR} a_{k\sigma R}^{\dagger }c_{\sigma }+({\rm H.c.})\}$
\hfil \break
and the direct transfer by \hfil \break
\indent $H_{LR}=\iint dkdq \{W_{kq}e^{i\varphi }a_{k\sigma L}^{\dagger }a_{q\sigma R}+({\rm H.c.})\}$, \hfil \break
where $W_{kq}=Wu_{kL}u_{qR}$ is assumed.
The phase $\varphi$ is proportional to the magnetic flux $\phi$ piercing between
the two paths: $\varphi = e\phi /(\hbar c)$ with $c$ the velocity of light.

\vspace*{-0.5cm}
\begin{figure}[htbp]
 \begin{center}
  \includegraphics[width=5cm,clip]{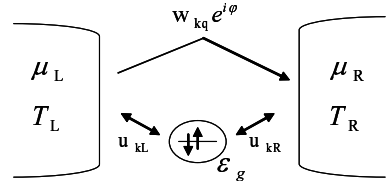}
 \end{center}
\vspace{-0.3cm}
\caption{Schematic view of an AB ring with a quantum dot}
\label{fig:sendaiquantum}
\end{figure}%
\vspace*{-0.5cm}

NESS can be simply characterized by
in-coming fields $\beta_{k\sigma \lambda }$. They are solutions of $[H,\beta_{k\sigma \lambda }]=-
\omega_{k\lambda } \beta_{k\sigma \lambda }$ with the boundary condition:
$a_{k\sigma \lambda }(t) \exp (i\omega _{k\lambda }t/\hbar ) \rightarrow 
\beta_{k\sigma \lambda }$ as $t\rightarrow -\infty $.
When $H$ does not have bound states,  
the original operators can be written as
\begin{align}
a_{k\sigma \lambda }=\beta _{k\sigma \lambda }
+\int &dk'\left(\dfrac{u_{k\lambda }u_{k'\lambda }}{\omega _{k'\lambda }-\omega _{k\lambda }+i0}
\dfrac{\chi _{k'\lambda }\beta _{k'\sigma \lambda }}{\Lambda(\omega_{k'\lambda })}\right. \notag\\
&+\left.\dfrac{u_{k\lambda }u_{k'\bar \lambda }}{\omega _{k'\bar \lambda }-
\omega _{k\lambda }+i0}\dfrac{\kappa _{k\bar \lambda }\beta _{k'\sigma \bar \lambda }}{\Lambda(\omega_{k'\bar \lambda })}\right)
\label{InComField1}
\end{align}
\begin{equation}
c_{\sigma }=\int dk\left(\dfrac{u_{k L}\Omega _{k L }\beta _{k\sigma L}}{\Lambda(\omega_{k L})}
+\dfrac{u_{k R }\Omega _{k R }\beta _{k\sigma R }}{\Lambda(\omega_{k R })}\right) \label{InComField2}
\end{equation}
where $\chi _{k\lambda }=1+M_{\lambda}(\omega_{k\lambda})[W^2(\omega _{k\lambda }-\epsilon _g)+2W\cos \varphi $] , \\
$\kappa_{k\lambda }=1+We^{i\varphi _{\lambda }}(\omega _{k\lambda }-\epsilon _{g})$,
$\Omega_{k\lambda }=1+We^{i\varphi _{\lambda }}M_{\lambda}(\omega_{k\lambda})$, \\
$\Lambda(x)=\nu(x) (x-\epsilon _g)-\sum_\lambda M _{\lambda}(x)-2W\cos\varphi M_{L}(x)M_{R}(x)$ \\
$\nu(x)=1-W^2M_{L}(x)M_{R}(x)$, and \\
$M_{\lambda}(x)=\displaystyle \int d\tilde k\frac{u_{\tilde k\lambda }^2}{x-\omega _{\tilde k\lambda }+i0}$.\\
In the above, ${\bar L}=R$, ${\bar R}=L$, $\varphi_L =-\varphi$ and $\varphi_R =\varphi$. 

NESS is constructed as follows: 
(1) Initially, the tunneling interactions are turned off and the two electrodes are
prepared to be in equilibrium with different temperatures ($T_L$ and  $T_R$) and 
chemical potentials ($\mu_L$ and $\mu_R$). The initial dot state can be taken arbitrarily.
Let $\rho_0$ be the density matrix describing this initial state.
\ \ 
(2) The tunneling interactions are turned on. Then, the whole state $\rho_t\equiv e^{-iHt/\hbar}\rho_0e^{iHt/\hbar}$ 
begins to evolve towards a steady state $\rho_+$. One can actually show that, when $t\to +\infty$, $\rho_t$ 
approaches to $\rho_+$ in a {\it weak} sense, namely, $\lim_{t\to +\infty}{\rm tr}(A \rho_t)={\rm tr}(A \rho_+)$
holds for every local observable (for detail, see Appendix). \ \ 

The so-obtained NESS $\rho_+$ satisfies Wick's theorem and
is characterized by the two-point function:
\begin{equation}
\langle \beta _{k\sigma \lambda }^{\dagger } \beta _{k'\sigma '\lambda '}\rangle =
F_{\lambda }(\epsilon )\delta (k-k'
)\delta _{\sigma \sigma '}\delta _{\lambda \lambda '}\label{eq:soukan}
\end{equation}  
where $\langle ...\rangle \equiv {\rm tr}(... \rho_+)$ stands for the NESS average and 
$F_{\lambda }(\epsilon )$ denotes the Fermi distribution 
function with temperature $T_\lambda$ and chemical potential $\mu_\lambda$ ($\lambda=L$ 
or $R$).

Here, a remark is in order. In spite of the fact that the system is conservative
(i.e., the evolution is governed by the Hamiltonian), the state evolution exhibits
`dissipation' in the {\it weak} sense. This behavior is due to the infinite extension
of the system. Indeed, information about the initial state is `dissipated' towards infinity 
(namely the Poincar\'e recurrence cycle is infinite) and cannot be
retrieved if we consider the average values of only the local observables.
Thus, for every local observable $A$, the long-term limit 
$\lim_{t\to +\infty}{\rm tr}(A \rho_t)$ exists.
Note that the Keldysh Green's function method shares the same property when applied to the
description of NESS.
In reality, the electrodes are finite, but can be regarded as infinite systems compared with
the mesoscopic subsystems (AB-ring and QD). And the experimentally observed nonequilibrium
states are considered to be well-described by NESS such as $\rho_+$.

Now we go back to the steady-state current. The operator $J$ corresponding to the
current from the left electrode is $J=-{d\over dt}\int dk\; a_{k\sigma L}^{\dagger }a_{k\sigma L}$
and its average is
\begin{equation}
\langle J \rangle =\displaystyle \frac{e}{h}\int dku_{kL}J_{k}^{\rm dot}\\ 
+\displaystyle \frac{e}{h}\iint dkdqWu_{kL}u_{qR}J_{kq}^{\rm arm} \label{denryu} 
\end{equation}
where $J_{k}^{\rm dot}=i\langle a_{k\sigma L}^{\dagger } c_{\sigma }\rangle +({\rm c.c.})$
is the current via the QD and $J_{kq}^{\rm arm}=i e^{i\varphi }\langle 
a_{k\sigma L}^{\dagger }a_{q\sigma R}\rangle +({\rm c.c.})$ is the current via the direct tunneling.
With the aid of eqs.(\ref{InComField1}), (\ref{InComField2}) and (\ref{eq:soukan}), the current 
flowing from the left electrode is given by
\begin{equation}
\langle J \rangle =\dfrac{e}{h}\; \displaystyle \int_{\mu_c}^{\infty}d\epsilon \; T(\epsilon )
\{F_L(\epsilon )-F_R(\epsilon )\}
\end{equation}
where $\mu_c =\max(\omega_{0L},\omega_{0R})$ and 
$T(\epsilon )$ is the transmission probability:
$$T(\epsilon )=\dfrac{4\pi ^2u^4D^2}{|\Lambda
(\epsilon )|^2}\{(W(\epsilon -\epsilon _g)+\cos \varphi )^2+\sin ^2\varphi\}.$$
The constant $D$ denotes the density of states of the two-dimensional electrodes.
For the sake of the simplicity, the tunneling matrix elements $u_{kL},u_{kR}$ are taken 
to be symmetric and independent of the wave number: \break $u_{kL}=u_{kR}=u$.
We remind that the bias voltage is equal to the chemical potential difference:
$\mu_L-\mu_R=eV$.
In the following sections, we focus on the nonequilibrium properties at zero temperature:
$T_L=T_R=0$.
\begin{figure}[b]
 \begin{center}
  \includegraphics[width=5.7cm,clip]{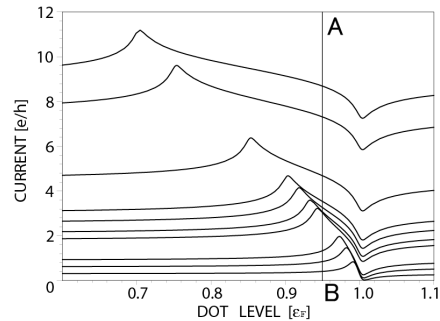}
 \end{center}
\vspace{-0.3cm}
\caption{Nonlinear current at zero temperature as a function of the gate voltage applied to the QD
for different bias voltages $eV$ from below $eV/\epsilon _{F}$=0.01, 0.02, 0.03, 0.06, 0.07, 0.085, 0.10, 0.15, 0.25, 0.30.
$\epsilon _{F}$ is the Fermi energy at $eV=0$.
The AB phase is fixed to $\varphi =0$ and $u^2DW=0.1$.
}
\label{fig:proceeding1}
\end{figure}

\section{Transport}
The dot-level-dependence of the current is presented in Fig.\ref{fig:proceeding1} for 
various bias voltages. 
In the linear response regime ($V\simeq 0$), when the dot level passes through the
Fermi energy of the electrodes, the resonant tunneling occurs and the current exhibits a
typical Fano peak.
When the bias voltage increases, the max-to-min distance becomes larger. 
This can be understood as follows.
The current peak is generated by the interference between electrons passing through
two conduction paths and this takes place when the dot energy $\epsilon_g$ 
lies between the two chemical potentials. Since the chemical potential difference 
is proportional to the bias voltage, the max-to-min distance of the current
increases with the bias voltage.
A similar deformation of the Fano peak as a function of the bias voltage was 
reported by Zhu {\it et al.}\cite{Yu}. 

\begin{figure}[t]
 \begin{center}
  \includegraphics[width=5.5cm,clip]{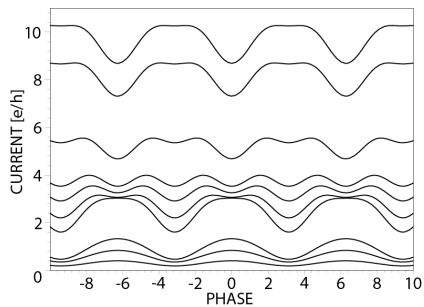}
 \end{center}
\vspace{-0.3cm}
\caption{Phase dependence of the current. The gate voltage is fixed to the A-B line shown in Fig.2. 
Other parameters are the same as those given in Fig.2.
}
\label{fig:proceeding2}
\end{figure}

The AB oscillations at a fixed dot level are depicted in Fig.\ref{fig:proceeding2}
for various bias voltages. The pattern of oscillation changes when one of the chemical
potentials passes through the dot level.
In particular, an oscillation with half wavelength appears when $|eV|=0.10\times \epsilon_F$
with $\epsilon_F$ the Fermi energy at zero bias. 

As is known\cite{Schuster}, the two-terminal conductance $G$ satisfies the Onsager
relation $G(\varphi )=G(-\varphi)$ and, thus, is an even function of $\varphi$.
This implies that the conductance takes its extremum at $\varphi=0$ or $\pi$.
Ueda {\it et al.}\cite{Ueda} pointed out that the conductance can take its maximum
at $\varphi \not=0, \pi$.
It is interesting to note that these two cases are simultaneously observed in
Fig.\ref{fig:proceeding2}.
When the bias voltage increases, the maximum of the current at
$\varphi=0$ splits into two maxima, which move to $\varphi=\pm\pi$ in a continuous way.
But the minimum of the current abruptly jumps from $\varphi=\pi$ to $\varphi=0$ at
the bias voltage where the oscillation with half wavelength appears. 
Similar AB oscillation patterns can be seen in the linear response regime by varying
the dot level\cite{Kang}. 

\begin{figure}[t]
 \begin{center}
  \includegraphics[width=5.5cm,clip]{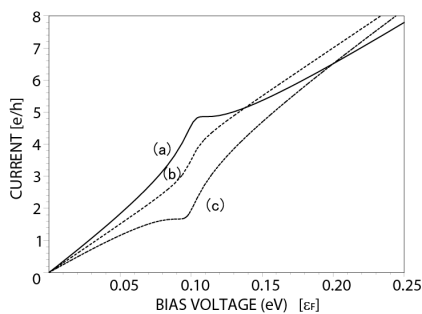}
 \end{center}
\vspace{-0.3cm}
\caption{Current versus bias voltage in units of $e/h$.
The phase affects the current-bias-voltage characteristics: the phase (a) $\varphi =0$ (the solid line),
(b) $\varphi= \pi /2$ (the dashed line) and 
(c) $\varphi =\pi $ (the dotted line).}
\label{fig:proceeding3}
\end{figure}

The strong bias-voltage-dependence of the AB oscillation suggests
a possibility of a device sensitive to the magnetic field.
Indeed, the current-bias-voltage characteristics
at a fixed dot level depend on $\varphi$ (Fig.\ref{fig:proceeding3}).

\section{Current Fluctuations}
Following Shimizu and Ueda\cite{Shimizu}, we investigate the fluctuation of the
charge $q(\tau)=\int^{\tau }_{0}dt J(t)$ leaving from the left electrode during the time interval 
$\tau$, where $J(t)$ is the current operator from the left electrode defined just before eq.(\ref{denryu}). 
Its fluctuation  
\begin{equation}
\Delta I\equiv \lim_{\tau \to \infty }\dfrac{\langle \{q(\tau)-\langle q(\tau)\rangle \}^2\rangle }{\tau }
\end{equation}
can be evaluated with the aid of Wick's theorem and the two-point function (\ref{eq:soukan})
and is given by
\begin{align}
\Delta I=\left(\frac{e}{2\pi h}\right)^2 \displaystyle &\int _{\mu _{c}}^{\infty }d\epsilon \Bigl\{T(\epsilon )
\sum_{\lambda } F_{\lambda }(\epsilon )\left[1-F_{\lambda }(\epsilon )\right] \notag \\
&+T(\epsilon )\left[1-T(\epsilon )\right]\left[F_{L}(\epsilon )-F_{R}(\epsilon )\right]^2\Bigr\}.
\end{align} 	 
The fluctuation $\Delta I$ consists of thermal and shot noises and, at zero temperature,
only the shot noise survives.

\begin{figure}[hbtp]
 \begin{center}
  \includegraphics[width=5cm,clip]{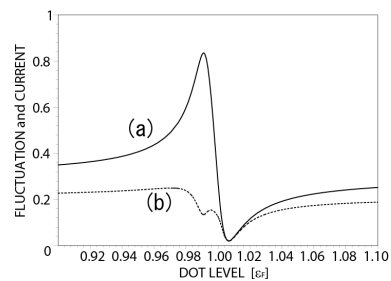}
 \end{center}
\vspace{-0.3cm}
\caption{The relationship between the current (a) and fluctuation (b) at
low bias voltage range ($eV/\epsilon _{F}=0.01$), $\varphi =0$, in units of
(a) $e/h$ and (b) $(e/2\pi h)^2$.}
\label{fig:proceeding4}
\end{figure}

Fig.\ref{fig:proceeding4} shows a typical behavior of $\Delta I$ as a function of the
dot level. One finds that $\Delta I$ has two dips, corresponding to the maximum and minimum 
of the current. Obviously, when the current is small, its fluctuation is small.
And thus, the current minimum corresponds to the fluctuation dip.
On the other hand, for one-dimensional perfect conductor, the current fluctuation is
known to vanish\cite{Shimizu} because all the available states are used for the
transport and there remains no room for fluctuations.
In the same way, when the current takes its maximum, the dot level is used for the
transport and is not available for
fluctuations. Then, the fluctuation is suppressed and the dip appears.
Similar results are reported in connection with the photo-assisted shot noise\cite{photon}.   

\begin{figure}[hbtp]
 \begin{center}
  \includegraphics[width=8.6cm,clip]{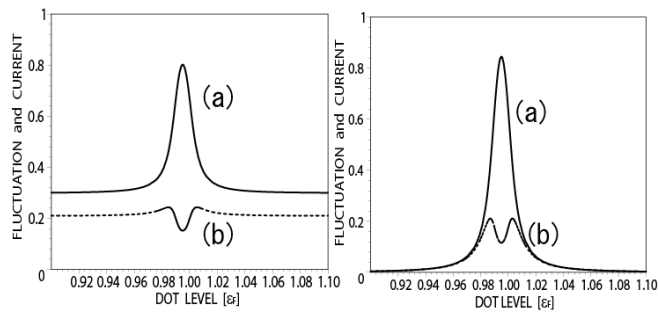}
 \end{center}
\vspace{-0.3cm}
\caption{The current (a) and fluctuation (b) with the present system at $\varphi =\pi /2$ (left) and
with the single dot system (right) 
in units of (a) $e/h$ and (b) $(e/2\pi h)^2$ 
at $eV/\epsilon _F=0.01$ respectively.
 }
\label{fig:proceeding5}
\end{figure}

When $\varphi=\pi/2$, the current is known to exhibit a symmetric Breit-Wigner-type 
resonance peak as for the system without direct tunneling between electrodes
(the single dot system). 
This seems to indicate the absence of interference between 
electrons passing through two paths. 
However, it is not the case and the interference can be seen in the current fluctuation.
Indeed, if the interference is absent, the fluctuation would be a sum of the fluctuation
of the single dot system and constant fluctuation of the system without quantum dot.
The curve (b) of the left figure of Fig.\ref{fig:proceeding5} depicts the fluctuation of the
present system at $\varphi=\pi/2$, clearly which is not a sum of a constant function
and the fluctuation of the single dot system (the curve (b) of the right figure of 
Fig.\ref{fig:proceeding5}). The difference can be attributed to the interference
between electrons passing through the two paths.

\vspace*{-1cm}

\vspace*{-0.5cm}
\section{Conclusions}
In summary, exact NESS for the noninteracting model of an AB ring with a QD
is explicitly constructed and bias-voltage-dependences
of the Fano resonance and the AB oscillation are investigated.
The AB oscillation patterns are shown to strongly depend on 
the bias voltage. This implies the nonlinear conduction sensitive to 
the magnetic field. 
The current fluctuations are studied as well and we show that, even when
the interference cannot be seen in the average current, it
contributes the current fluctuation.

\section*{Acknowledgment}

The authors thank Professor T. Matsui for discussions.
One of them (JT) is grateful to Professor S. Tarucha
for fruitful discussions and valuable comments. This work is 
partially supported 
by a Grant-in-Aid for Scientific 
Research of Priority Areas ``Control of Molecules in Intense Laser Fields'' 
and the 21st Century COE Program at Waseda University ``Holistic Research and 
Education Center for Physics of Self-organization Systems''
both from the Ministry of Education, Culture, Sports, Science and 
Technology of Japan.

\appendix
\section{C$^*$-algebraic approach to NESS}

In this approach\cite{Bratteli}, one starts from a set of all (spatially) local observables, 
which form an algebra called the C$^*$-algebra. 
Remind that, if the interaction range is finite, the commutator $i[H,A]$ 
of the total Hamiltonian $H$ and a local observable $A$ is a local
operator. Thus, both sides of the Heisenberg equation of motion
$\hbar{\partial A\over \partial t}=i[H,A]$ are local and the time evolution
$A(t)$ can be treated within the framework of local observables.

Physical states $\rho$ are introduced by specifying average 
values $\langle A\rangle_\rho$ of all local observables $A$ (i.e., the 
states are (positive) linear functionals over the algebra of local observables).
Then, the long-term behavior of the system is naturally studied through
the behaviors of the averages $\langle A\rangle_{\rho(t)}$.
Then, for various systems, one can show the existence of the weak limits
$\rho_\pm$ defined by $\lim_{t\to\pm \infty}\langle A\rangle_{\rho(t)}
=\langle A\rangle_{\rho_\pm}$ for any observable $A$.
In this way, one can avoid the explicit use of infinite quantities such as the total 
Hamiltonian and total particle numbers, which are mathematically ill-defined, and
one may realize the unidirectional state evolution in the weak sense.

As mentioned in the text, our model is a typical example of the junction systems 
studied by Fr\"olich {\it et al.}\cite{Frolich}. Since they rigorously show the existence 
of NESS and give its mathematical characterizations, we give a heuristic proof of the existence of NESS 
and its characterization presented in the text.

Let $H_0 \equiv H_L+H_R$ and define the M\o ller operator \break
$\Omega ={\displaystyle \lim_{t\to -\infty}} e^{iHt/\hbar}e^{-iH_0t/\hbar}=
{\displaystyle \lim_{t\to +\infty}} e^{-iHt/\hbar}e^{iH_0t/\hbar}$,
then the incoming field operators are expressed as \hfil \break
$\beta_{k\sigma \lambda}\mskip -5mu =\mskip -5mu {\displaystyle 
\lim_{t\to -\infty}}e^{iHt/\hbar}\mskip -5mu e^{-iH_0t/\hbar}
a_{k\sigma \lambda}e^{iH_0t/\hbar}\mskip -5mu e^{-iHt/\hbar}\mskip -5mu =\mskip -5mu 
\Omega a_{k\sigma \lambda}\Omega^\dag$.
Suppose that the total Hamiltonian $H$ does not admit bound states, then, since $\Omega$ intertwines $H$ with 
$H_0$: $H\Omega=\Omega H_0$ and $H_0$ has bound states represented by $c_\sigma$, $\Omega$ should annihilate
the states involving dot electrons and, thus, the product $\Pi\equiv \Omega^\dag \Omega$ is a
projection onto the states consisting only the electrons represented by $a_{k\sigma \lambda}$ (see the arguments in 
a standard text book on scattering theory\cite{Newton}). 
Therefore, one has \break
$\Pi=(1-c_\uparrow^\dag c_\uparrow)(1-c_\downarrow^\dag c_\downarrow)$
and it commutes with $a_{k\sigma \lambda}$.  
On the other hand,
remind that the initial state can be written as 
$\rho_0\propto \rho_c \exp \left[-\sum_\lambda (H_\lambda-\mu_\lambda N_\lambda)/T_\lambda \right]$ where 
$N_\lambda$ is the total number operator of each electrode and $\rho_c$ is
the initial dot state. Since $\rho_0$ commutes with $H_0$, the steady state is given by \break
$\rho_+={\displaystyle \lim_{t\to +\infty}} e^{-iHt/\hbar}e^{iH_0t/\hbar} 
\rho_0 e^{-iH_0t/\hbar}e^{iHt/\hbar}$. This seems to imply $\rho_+=\Omega \rho_0 \Omega^\dag$,
but the normalization changes as in the wave function renormalization, and one has 
$\rho_+=\Omega \rho_0 \Omega^\dag/{\rm tr}(\Omega \rho_0 \Omega^\dag)=\Omega \rho_0 \Omega^\dag/\langle\Pi\rangle_0$,
where $\langle \cdots \rangle_0$ stands for ${\rm tr}(\cdots \rho_0)$.
Hence, for example,
\begin{eqnarray}
&&\langle \beta_{k_1\sigma_1 \lambda_1}^\dag\beta_{k_2\sigma_2 \lambda_2}^\dag \beta_{k_3\sigma_3 \lambda_3}
\beta_{k_4\sigma_4 \lambda_4}\rangle \nonumber \\
&&={\langle \Pi a_{k_1\sigma_1 \lambda_1}^\dag \Pi a_{k_2\sigma_2 \lambda_2}^\dag 
\Pi a_{k_3\sigma_3 \lambda_3}\Pi 
a_{k_4\sigma_4 \lambda_4}\Pi  \rangle_0
\over \langle \Pi \rangle_0}
\nonumber \\
&&={\langle a_{k_1\sigma_1 \lambda_1}^\dag a_{k_2\sigma_2 \lambda_2}^\dag 
a_{k_3\sigma_3 \lambda_3} 
a_{k_4\sigma_4 \lambda_4}\Pi  \rangle_0/
\langle \Pi \rangle_0}
\nonumber \\
&&=\langle a_{k_1\sigma_1 \lambda_1}^\dag a_{k_2\sigma_2 \lambda_2}^\dag 
a_{k_3\sigma_3 \lambda_3} a_{k_4\sigma_4 \lambda_4} \rangle_0 \ . \nonumber
\end{eqnarray}
The last equality follows from the fact that $a_{k\sigma \lambda}$ and $c_\sigma$ (hence, $\Pi$) 
are statistically independent at the state $\rho_0$.
Therefore, since the initial state $\rho_0$ satisfies Wick's theorem for $a_{k\sigma \lambda}$,
the same is true for $\rho_+$ and $\beta_{k\sigma \lambda}$ and
the two-point functions
$\langle \beta_{k\sigma \lambda}^\dag\beta_{k'\sigma' \lambda'}\rangle$ are given by (\ref{eq:soukan}).


\begin{thebibliography}{99} 

\bibitem{Iye} K. Kobayashi, H. Aikawa, S. Katsumoto and Y. Iye: Phys. Rev. Lett. \textbf{88} 256806 (2002).
\bibitem{FANO} U. Fano: Phys. Rev. \textbf{124} 1866 (1961).
\bibitem{Yacoby} A. Yacoby, M. Heiblum, D. Mahalu and H. Shtrikman: Phys. Rev. Lett. \textbf{74} 4047 (1995).

\bibitem{Ojima} I. Ojima, H. Hasegawa and M.
Ichiyanagi: J. Stat. Phys. {\bf 50}, 633
(1988); I. Ojima: J. Stat. Phys. {\bf 56},
203 (1989); in {\em Quantum Aspects of Optical
Communications}, (LNP {\bf 378},Springer,1991).

\bibitem{Ruelle} D. Ruelle: J. Stat. Phys.  {\bf 98}, 57 (2000);
Comm. Math. Phys. {\bf 224 }, 3 (2001).

\bibitem{HoAraki} T.G. Ho and H. Araki:
{\em Proc. Steklov Math. Institute} {\bf 228}, (2000) 191.

\bibitem{Jaksic} V. Jak\v si\'c and C.-A. Pillet: Commun. Math. Phys. {\bf 217}, 285 (2001);
Commun. Math. Phys. {\bf 226}, 131 (2002); J. Stat. Phys. {\bf 108}, 269 (2002)
and references therein.

\bibitem{ST} S. Tasaki: Chaos, Solitons and Fractals, {\bf 12} 2657 (2001); 
S. Tasaki, T. Matsui: in {\it Fundamental Aspects of Quantum Physics} 
eds. L. Accardi, S. Tasaki: World Scientific, Singapore, (2003) p.100
and references therein.

\bibitem{Frolich} J. Fr\"ohlich, M. Merkli, D. Ueltschi:
{\it Ann. Henri Poincar\'e} {\bf 4}, 897 (2003).

\bibitem{Bratteli}  O. Bratteli and D.W. Robinson:
{\em Operator Algebras and Quantum
Statistical Mechanics} vol.1 (Springer, New York,
1987); vol.2, (Springer, New York,
1997).

\bibitem{RuelleEq} D. Ruelle: {\it Statistical Mechanics:
Rigorous Results}, (Benjamin, Reading, 1969); Ya. G. Sinai: {\it
The Theory of Phase Transitions: Rigorous Results}, (Pergamon,
Oxford, 1982).

\bibitem{Walter} W. Hofstetter, J. K\"{o}nig and H. Schoeller: Phys. Rev. Lett. \textbf{87} 156803 (2001).

\bibitem{Yu} Y. Zhu, Q. Sun and T. Lin: cond-mat/02029456.

\bibitem{Schuster} R. Schuster, E. Buks, M. Heiblum, D. Mahalu, V. Umansky and H. Shtrikman:
Nature \textbf{385} 417 (1997).

\bibitem{Ueda} A. Ueda, I. Baba, K. Suzuki and M. Eto: {\em Proc. Int. Conf. on Quantum Transport and Quantum Coherence, 
Tokyo, 2002},
J. Phys. Soc. Jpn \textbf{72} (2003) Suppl.A, p.157-158.

\bibitem{Kang} K. Kang: Phys. Rev. \textbf{59} 4608 (1999).
\bibitem{Shimizu}A. Shimizu, M. Ueda: Phys. Rev. Lett. \textbf{69} 1403 (1992).
\bibitem{photon}Z. Ma, Y. Zhu, X. Li, T. Lin and Z.Su: Phys. Rev. B \textbf{69} 045302 (2004). 

\bibitem{Newton} For example, R.G. Newton:{\it Scattering Theory of Waves and Particles},
(McGraw-Hill, New York, 1966).


\end{thebibliography}
\end{document}